\documentclass[twocolumn]{aastex631}
\usepackage{color}
\usepackage{xspace}
\usepackage{graphicx}
\usepackage{amsmath}
\usepackage{amssymb}
\usepackage{yfonts}
\usepackage{xcolor}
\usepackage{float}
\hypersetup{pdfauthor={Name}}
\usepackage{stackengine}
\shorttitle{Neutrinos from GRB 221009A}
\shortauthors{Veres et al.}

\begin{document}

\title{Non-detection of Neutrinos from the BOAT:\\Improved Constraints on the Parameters of GRB~221009A }

\correspondingauthor{P. Veres}
\email{peter.veres@uah.edu} 

%%%%%%%%%%%%%%%%%%%%%%%%%%%%%%%%%%%%%%%
%%                                    %%
%%            Definitions             %%
%%                                    %%
%%%%%%%%%%%%%%%%%%%%%%%%%%%%%%%%%%%%%%%%

%%%%%%%%%%%%%%%%%%%%%%%%%%%%%%%%%%%%%%%%
%%                Math                %%
%%%%%%%%%%%%%%%%%%%%%%%%%%%%%%%%%%%%%%%%

%% make tensor symbol %%
\def\shrinkage{2.1mu}
\def\vecsign{\mathchar"017E}
\def\dvecsign{\smash{\stackon[-1.95pt]{\mkern-\shrinkage\vecsign}{\rotatebox{180}{$\mkern-\shrinkage\vecsign$}}}}
\def\dblvec#1{\def\useanchorwidth{T}\stackon[-4.2pt]{#1}{\,\dvecsign}}
\stackMath
%% uppercase letters %%
% vectors
\def\Avec{\vec{A}}
\def\Bvec{\vec{B}}
\def\Cvec{\vec{C}}
\def\Dvec{\vec{D}}
\def\Evec{\vec{E}}
\def\Fvec{\vec{F}}
\def\Gvec{\vec{G}}
\def\Hvec{\vec{H}}
\def\Ivec{\vec{I}}
\def\Jvec{\vec{J}}
\def\Kvec{\vec{K}}
\def\Lvec{\vec{L}}
\def\Mvec{\vec{M}}
\def\Nvec{\vec{N}}
\def\Ovec{\vec{O}}
\def\Pvec{\vec{P}}
\def\Qvec{\vec{Q}}
\def\Rvec{\vec{R}}
\def\Svec{\vec{S}}
\def\Tvec{\vec{T}}
\def\Uvec{\vec{U}}
\def\Vvec{\vec{V}}
\def\Wvec{\vec{W}}
\def\Xvec{\vec{X}}
\def\Yvec{\vec{Y}}
\def\Zvec{\vec{Z}}
% unit vectors
\def\Ahat{\hat{A}}
\def\Bhat{\hat{B}}
\def\Chat{\hat{C}}
\def\Dhat{\hat{D}}
\def\Ehat{\hat{E}}
\def\Fhat{\hat{F}}
\def\Ghat{\hat{G}}
\def\Hhat{\hat{H}}
\def\Ihat{\hat{I}}
\def\Jhat{\hat{J}}
\def\Khat{\hat{K}}
\def\Lhat{\hat{L}}
\def\Mhat{\hat{M}}
\def\Nhat{\hat{N}}
\def\Ohat{\hat{O}}
\def\Phat{\hat{P}}
\def\Qhat{\hat{Q}}
\def\Rhat{\hat{R}}
\def\Shat{\hat{S}}
\def\That{\hat{T}}
\def\Uhat{\hat{U}}
\def\Vhat{\hat{V}}
\def\What{\hat{W}}
\def\Xhat{\hat{X}}
\def\Yhat{\hat{Y}}
\def\Zhat{\hat{Z}}
% tensors
\def\Aten{\dblvec{A}}
\def\Bten{\dblvec{B}}
\def\Cten{\dblvec{C}}
\def\Dten{\dblvec{D}}
\def\Eten{\dblvec{E}}
\def\Ften{\dblvec{F}}
\def\Gten{\dblvec{G}}
\def\Hten{\dblvec{H}}
\def\Iten{\dblvec{I}}
\def\Jten{\dblvec{J}}
\def\Kten{\dblvec{K}}
\def\Lten{\dblvec{L}}
\def\Mten{\dblvec{M}}
\def\Nten{\dblvec{N}}
\def\Oten{\dblvec{O}}
\def\Pten{\dblvec{P}}
\def\Qten{\dblvec{Q}}
\def\Rten{\dblvec{R}}
\def\Sten{\dblvec{S}}
\def\Tten{\dblvec{T}}
\def\Uten{\dblvec{U}}
\def\Vten{\dblvec{V}}
\def\Wten{\dblvec{W}}
\def\Xten{\dblvec{X}}
\def\Yten{\dblvec{Y}}
\def\Zten{\dblvec{Z}}
% outlined
\def\Aol{$\mathbbm{A}$}
\def\Bol{$\mathbbm{B}$}
\def\Col{$\mathbbm{C}$}
\def\Dol{$\mathbbm{D}$}
\def\Eol{$\mathbbm{E}$}
\def\Fol{$\mathbbm{F}$}
\def\Gol{$\mathbbm{G}$}
\def\Hol{$\mathbbm{H}$}
\def\Iol{$\mathbbm{I}$}
\def\Jol{$\mathbbm{J}$}
\def\Kol{$\mathbbm{K}$}
\def\Lol{$\mathbbm{L}$}
\def\Mol{$\mathbbm{M}$}
\def\Nol{$\mathbbm{N}$}
\def\Ool{$\mathbbm{O}$}
\def\Pol{$\mathbbm{P}$}
\def\Qol{$\mathbbm{Q}$}
\def\Rol{$\mathbbm{R}$}
\def\Sol{$\mathbbm{S}$}
\def\Tol{$\mathbbm{T}$}
\def\Uol{$\mathbbm{U}$}
\def\Vol{$\mathbbm{V}$}
\def\Wol{$\mathbbm{W}$}
\def\Xol{$\mathbbm{X}$}
\def\Yol{$\mathbbm{Y}$}
\def\Zol{$\mathbbm{Z}$}
%% lowercase letters %%
% vectors
\def\avec{\vec{a}}
\def\bvec{\vec{b}}
\def\cvec{\vec{c}}
\def\dvec{\vec{d}}
\def\evec{\vec{e}}
\def\fvec{\vec{f}}
\def\gvec{\vec{g}}
\def\hvec{\vec{h}}
\def\ivec{\vec{i}}
\def\jvec{\vec{j}}
\def\kvec{\vec{k}}
\def\lvec{\vec{l}}
\def\mvec{\vec{m}}
\def\nvec{\vec{n}}
\def\ovec{\vec{o}}
\def\pvec{\vec{p}}
\def\qvec{\vec{q}}
\def\rvec{\vec{r}}
\def\svec{\vec{s}}
\def\tvec{\vec{t}}
\def\uvec{\vec{u}}
\def\vvec{\vec{v}}
\def\wvec{\vec{w}}
\def\xvec{\vec{x}}
\def\yvec{\vec{y}}
\def\zvec{\vec{z}}
% unit vectors
\def\ahat{\hat{a}}
\def\bhat{\hat{b}}
\def\chat{\hat{c}}
\def\dhat{\hat{d}}
\def\ehat{\hat{e}}
\def\fhat{\hat{f}}
\def\ghat{\hat{g}}
\def\hhat{\hat{h}}
\def\ihat{\hat{i}}
\def\jhat{\hat{j}}
\def\khat{\hat{k}}
\def\lhat{\hat{l}}
\def\mhat{\hat{m}}
\def\nhat{\hat{n}}
\def\ohat{\hat{o}}
\def\phat{\hat{p}}
\def\qhat{\hat{q}}
\def\rhat{\hat{r}}
\def\shat{\hat{s}}
\def\that{\hat{t}}
\def\uhat{\hat{u}}
\def\vhat{\hat{v}}
\def\what{\hat{w}}
\def\xhat{\hat{x}}
\def\yhat{\hat{y}}
\def\zhat{\hat{z}}
% tensors
\def\aten{\dblvec{a}}
\def\bten{\dblvec{b}}
\def\cten{\dblvec{c}}
\def\dten{\dblvec{d}}
\def\eten{\dblvec{e}}
\def\ften{\dblvec{f}}
\def\gten{\dblvec{g}}
\def\hten{\dblvec{h}}
\def\iten{\dblvec{i}}
\def\jten{\dblvec{j}}
\def\kten{\dblvec{k}}
\def\lten{\dblvec{l}}
\def\mten{\dblvec{m}}
\def\nten{\dblvec{n}}
\def\oten{\dblvec{o}}
\def\pten{\dblvec{p}}
\def\qten{\dblvec{q}}
\def\rten{\dblvec{r}}
\def\sten{\dblvec{s}}
\def\tten{\dblvec{t}}
\def\uten{\dblvec{u}}
\def\vten{\dblvec{v}}
\def\wten{\dblvec{w}}
\def\xten{\dblvec{x}}
\def\yten{\dblvec{y}}
\def\zten{\dblvec{z}}
% outlined
\def\aol{$\mathbbm{a}$}
\def\bol{$\mathbbm{b}$}
\def\col{$\mathbbm{c}$}
\def\dol{$\mathbbm{d}$}
\def\eol{$\mathbbm{e}$}
\def\fol{$\mathbbm{f}$}
\def\gol{$\mathbbm{g}$}
\def\hol{$\mathbbm{h}$}
\def\iol{$\mathbbm{i}$}
\def\jol{$\mathbbm{j}$}
\def\kol{$\mathbbm{k}$}
\def\lol{$\mathbbm{l}$}
\def\mol{$\mathbbm{m}$}
\def\nol{$\mathbbm{n}$}
\def\ool{$\mathbbm{o}$}
\def\pol{$\mathbbm{p}$}
\def\qol{$\mathbbm{q}$}
\def\rol{$\mathbbm{r}$}
\def\sol{$\mathbbm{s}$}
\def\tol{$\mathbbm{t}$}
\def\uol{$\mathbbm{u}$}
\def\vol{$\mathbbm{v}$}
\def\wol{$\mathbbm{w}$}
\def\xol{$\mathbbm{x}$}
\def\yol{$\mathbbm{y}$}
\def\zol{$\mathbbm{z}$}
%% greek letters %%
% renaming
\newcommand{\eps}{\epsilon}
\newcommand{\veps}{\varepsilon}
\newcommand{\ve}{\varepsilon}
\newcommand{\vtheta}{\vartheta}
\newcommand{\vphi}{\varphi}
\newcommand{\vrho}{\varrho}
% vectors
\def\alphavec{\vec{\alpha}}
\def\nuvec{\vec{\nu}}
\def\betavec{\vec{\beta}}
\def\xivec{\vec{\xi}}
\def\Xivec{\vec{\Xi}}
\def\gammavec{\vec{\gamma}} 
\def\Gammavec{\vec{\Gamma}}
\def\deltavec{\vec{\delta}} 
\def\Deltavec{\vec{\Delta}}
\def\pivec{\vec{\pi}} 
\def\Pivec{\vec{\Pi}}
\def\epsvec{\vec{\eps}} 
\def\vepsvec{\vec{\veps}} 
\def\rhovec{\vec{\rho}}
\def\vrhovec{\vec{\vrho}}
\def\zetavec{\vec{\zeta}}
\def\sigmavec{\vec{\sigma}}
\def\Sigmavec{\vec{\Sigma}}
\def\etavec{\vec{\eta}}
\def\tauvec{\vec{\tau}}
\def\thetavec{\vec{\theta}}
\def\vthetavec{\vec{\vtheta}}
\def\Thetavec{\vec{\Theta}}
\def\upsilonvec{\vec{\upsilon}}
\def\Upsilonvec{\vec{\Upsilon}}
\def\iotavec{\vec{\iota}}
\def\phivec{\vec{\phi}}
\def\vphivec{\vec{\vphi}}
\def\Phivec{\vec{\Phi}}
\def\kappavec{\vec{\kappa}}
\def\chivec{\vec{\chi}}
\def\lambdavec{\vec{\lambda}}
\def\Lambdavec{\vec{\Lambda}}
\def\psivec{\vec{\psi}}
\def\Psivec{\vec{\Psi}}
\def\muvec{\vec{\mu}}
\def\omegavec{\vec{\omega}}
\def\Omegavec{\vec{\Omega}}
% unit vectors
\def\alphahat{\hat{\alpha}}
\def\nuhat{\hat{\nu}}
\def\betahat{\hat{\beta}}
\def\xihat{\hat{\xi}}
\def\Xihat{\hat{\Xi}}
\def\gammahat{\hat{\gamma}} 
\def\Gammahat{\hat{\Gamma}}
\def\deltahat{\hat{\delta}} 
\def\Deltahat{\hat{\Delta}}
\def\pihat{\hat{\pi}} 
\def\Pihat{\hat{\Pi}}
\def\epshat{\hat{\eps}} 
\def\vepshat{\hat{\veps}} 
\def\rhohat{\hat{\rho}}
\def\vrhohat{\hat{\vrho}}
\def\zetahat{\hat{\zeta}}
\def\sigmahat{\hat{\sigma}}
\def\Sigmahat{\hat{\Sigma}}
\def\etahat{\hat{\eta}}
\def\tauhat{\hat{\tau}}
\def\thetahat{\hat{\theta}}
\def\vthetahat{\hat{\vtheta}}
\def\Thetahat{\hat{\Theta}}
\def\upsilonhat{\hat{\upsilon}}
\def\Upsilonhat{\hat{\Upsilon}}
\def\iotahat{\hat{\iota}}
\def\phihat{\hat{\phi}}
\def\vphihat{\hat{\vphi}}
\def\Phihat{\hat{\Phi}}
\def\kappahat{\hat{\kappa}}
\def\chihat{\hat{\chi}}
\def\lambdahat{\hat{\lambda}}
\def\Lambdahat{\hat{\Lambda}}
\def\psihat{\hat{\psi}}
\def\Psihat{\hat{\Psi}}
\def\muhat{\hat{\mu}}
\def\omegahat{\hat{\omega}}
\def\Omegahat{\hat{\Omega}}
% tensors
\def\alphaten{\dblvec{\alpha}}
\def\nuten{\dblvec{\nu}}
\def\betaten{\dblvec{\beta}}
\def\xiten{\dblvec{\xi}}
\def\Xiten{\dblvec{\Xi}}
\def\gammaten{\dblvec{\gamma}} 
\def\Gammaten{\dblvec{\Gamma}}
\def\deltaten{\dblvec{\delta}} 
\def\Deltaten{\dblvec{\Delta}}
\def\piten{\dblvec{\pi}} 
\def\Piten{\dblvec{\Pi}}
\def\epsten{\dblvec{\eps}} 
\def\vepsten{\dblvec{\veps}} 
\def\rhoten{\dblvec{\rho}}
\def\vrhoten{\dblvec{\vrho}}
\def\zetaten{\dblvec{\zeta}}
\def\sigmaten{\dblvec{\sigma}}
\def\Sigmaten{\dblvec{\Sigma}}
\def\etaten{\dblvec{\eta}}
\def\tauten{\dblvec{\tau}}
\def\thetaten{\dblvec{\theta}}
\def\vthetaten{\dblvec{\vtheta}}
\def\Thetaten{\dblvec{\Theta}}
\def\upsilonten{\dblvec{\upsilon}}
\def\Upsilonten{\dblvec{\Upsilon}}
\def\iotaten{\dblvec{\iota}}
\def\phiten{\dblvec{\phi}}
\def\vphiten{\dblvec{\vphi}}
\def\Phiten{\dblvec{\Phi}}
\def\kappaten{\dblvec{\kappa}}
\def\chiten{\dblvec{\chi}}
\def\lambdaten{\dblvec{\lambda}}
\def\Lambdaten{\dblvec{\Lambda}}
\def\psiten{\dblvec{\psi}}
\def\Psiten{\dblvec{\Psi}}
\def\muten{\dblvec{\mu}}
\def\omegaten{\dblvec{\omega}}
\def\Omegaten{\dblvec{\Omega}}
% outlined
\def\alphaol{$\mathbb{\alpha}$}
\def\nuol{$\mathbb{\nu}$}
\def\betaol{$\mathbb{\beta}$}
\def\xiol{$\mathbb{\xi}$}
\def\Xiol{$\mathbb{\Xi}$}
\def\gammaol{$\mathbb{\gamma}$}
\def\Gammaol{$\mathbb{\Gamma}$}
\def\deltaol{$\mathbb{\delta}$}
\def\Deltaol{$\mathbb{\Delta}$}
\def\piol{$\mathbb{\pi}$}
\def\Piol{$\mathbb{\Pi}$}
\def\epsol{$\mathbb{\eps}$}
\def\vepsol{$\mathbb{\veps}$}
\def\rhool{$\mathbb{\rho}$}
\def\vrhool{$\mathbb{\vrho}$}
\def\zetaol{$\mathbb{\zeta}$}
\def\sigmaol{$\mathbb{\sigma}$}
\def\Sigmaol{$\mathbb{\Sigma}$}
\def\etaol{$\mathbb{\eta}$}
\def\tauol{$\mathbb{\tau}$}
\def\thetaol{$\mathbb{\theta}$}
\def\vthetaol{$\mathbb{\vtheta}$}
\def\Thetaol{$\mathbb{\Theta}$}
\def\upsilonol{$\mathbb{\upsilon}$}
\def\Upsilonol{$\mathbb{\Upsilon}$}
\def\iotaol{$\mathbb{\iota}$}
\def\phiol{$\mathbb{\phi}$}
\def\vphiol{$\mathbb{\vphi}$}
\def\Phiol{$\mathbb{\Phi}$}
\def\kappaol{$\mathbb{\kappa}$}
\def\chiol{$\mathbb{\chi}$}
\def\lambdaol{$\mathbb{\lambda}$}
\def\Lambdaol{$\mathbb{\Lambda}$}
\def\psiol{$\mathbb{\psi}$}
\def\Psiol{$\mathbb{\Psi}$}
\def\muol{$\mathbb{\mu}$}
\def\omegaol{$\mathbb{\omega}$}
\def\Omegaol{$\mathbb{\Omega}$}
%%%%%%%%%%%%%%%%%%%%%%%%%%%%%%
% vector calculus
\def\cross{\times}
\def\del{\nabla}
\def\delcross{\nabla \times}
\def\deldot{\nabla \cdot}
\def\delsq{\nabla^2}
% arrows
\newcommand{\rarrow}{\Rightarrow}
\newcommand{\rrarrow}{\Longrightarrow}
\newcommand{\larrow}{\Leftarrow}
\newcommand{\llarrow}{\Longleftarrow}
\newcommand{\lrarrow}{\Leftrightarrow}
\newcommand{\llrrarrow}{\iff}
% geometry operators
\newcommand{\nperp}{\not\perp}
% infinity
\let\oldinf\inf
\renewcommand{\inf}{\infty}
% boxes
\def\wbox{\square}
\def\bbox{\blacksquare}
% degrees
\def\deg{^{\circ}}
% plain text
\newcommand{\trm}{\textrm}
\newcommand{\tbf}{\textbf}
\newcommand{\tul}{\underline}
\newcommand{\tit}{\textit}
\newcommand{\texp}[1]{$^{\textrm{#1}}$}%texponent
\newcommand{\tqu}{\enquote}%quotes
\newcommand{\pref}[1]{(\pageref{#1})}
\newcommand{\eref}[1]{equation \eqref{#1}}
% average
\newcommand{\avg}[1]{\overline{#1}}
% ()
\newcommand{\p}[1]{\left( #1 \right)}
% []
\newcommand{\pp}[1]{\left[ #1 \right]}
% {}
\newcommand{\psqu}[1]{\left\{ #1 \right\}}
% <>
\newcommand{\pang}[1]{\left\langle #1 \right\rangle}
% ||
\newcommand{\abs}[1]{\left| #1 \right|}
% ||||
\newcommand{\dabs}[1]{\left\lVert #1 \right\rVert}
% integral evaluation
\newcommand{\eval}[2]{\rvert_{#1}^{#2}}
\newcommand{\Eval}[2]{\Bigg\rvert_{#1}^{#2}}
% scientific notation
\newcommand{\e}[1]{\times 10^{#1}}
% derivatives
\newcommand{\dv}[2]{\frac{d #1}{d #2}}
\newcommand{\ndv}[3]{\frac{d^{#1} #2}{d #3^{#1}}}
\newcommand{\pdv}[2]{\frac{\partial #1}{\partial #2}}
\newcommand{\npdv}[3]{\frac{\partial^{#1} #2}{\partial #3^{#1}}}
% math labels
\newcommand{\ulabel}[2]{\underset{\mathclap{\substack{\uparrow\\#2}}}{#1}}
\newcommand{\llabel}[2]{\overset{\mathclap{\substack{#2\\\downarrow}}}{#1}}
\newcommand{\ublabel}[2]{\overbrace{#1}^{\mathclap{\substack{#2}}}}
\newcommand{\lblabel}[2]{\underbrace{#1}_{\mathclap{\substack{#2}}}}
\newcommand{\uslabel}[2]{\overbracket{#1}^{\mathclap{\substack{#2}}}}
\newcommand{\lslabel}[2]{\underbracket{#1}_{\mathclap{\substack{#2}}}}
% limit
\let\oldlim\lim
\renewcommand{\lim}[2]{\oldlim\limits_{{#1} \rightarrow {#2}}}
%sum
\let\oldsum\sum
\renewcommand{\sum}[2]{\oldsum\limits_{#1}^{#2}}
% product
\let\oldprod\prod
\renewcommand{\prod}[2]{\oldprod\limits_{#1}^{#2}}
%% integral
% equation
\newcommand{\eq}[1]{\begin{equation*} #1 \end{equation*}}
% equation labeled
\newcommand{\eql}[2]{\begin{equation} \label{#1} #2 \end{equation}}
% matrix
\let\oldmatrix\matrix
\renewcommand{\matrix}[1]{$\begin{pmatrix} #1 \end{pmatrix}$}

%%%%%%%%%%%%%%%%%%%%%%%%%%%%%%%%%%%%%%%%
%%       Physics & Astrophysics       %%
%%%%%%%%%%%%%%%%%%%%%%%%%%%%%%%%%%%%%%%%

% length
\def\nm{\mbox{~nm}} % 10^-9
\def\mum{\mbox{~\mu\hbox{m}}} % 10^-6
\def\mm{\mbox{~mm}} % 10^-3
\def\cm{\mbox{~cm}} % 10^-2
\def\m{\mbox{~m}} % 10^0
\def\km{\mbox{~km}} % 10^3
% distance
\def\pc{\mbox{~pc}} % 10^0
\def\kpc{\mbox{~kpc}} % 10^3
\def\Mpc{\mbox{~Mpc}} % 10^6
\def\Gpc{\mbox{~Gpc}} % 10^9
% energy
\def\erg{\mbox{~erg}}
\def\eV{\mbox{~eV}} % 10^0
\def\keV{\mbox{~keV}} % 10^3
\def\MeV{\mbox{~MeV}} % 10^6
\def\GeV{\mbox{~GeV}} % 10^9
\def\TeV{\mbox{~TeV}} % 10^12
% frequency
\def\Hz{\mbox{~Hz}} % 10^0
% time
\def\ns{\mbox{~ns}} % 10^-9
\def\mus{~\mu\hbox{s}} % 10^-6
\def\ms{\mbox{~ms}} % 10^-3
\def\sec{\mbox{~s}} % 10^0
\def\s{\mbox{~s}} % 10^0
\def\hr{\mbox{~h}}
\def\yr{\mbox{~yr}}
\def\Jy{\mbox{~Jy}}
% velocity
\def\ms{{\mbox{~ms}}}
% astro constants
\def\astar{A$_{\star}$}
\def\Msun{M$_{\odot}$}
% GRBs
\def\Epk{E$_{\textrm{peak}}$}
\def\Eiso{E$_{\textrm{iso}}$}
\def\Liso{L$_{\textrm{iso}}$}
\def\t90{T$_{\textrm{90}}$}
\def\tvar{t_{\textrm{var}}}
\def\ra#1#2#3{#1$^{^\textrm{h}}$#2$^{^\textrm{m}}$#3$^{^\textrm{s}}$}
\def\dec#1#2#3{#1$^\circ$#2$'$#3$''$}
\def\swift{{\textit{Swift}}}
\def\fermi{{\textit{Fermi}}}

%%%%%%%%%%%%%%%%%%%%%%%%%%%%%%%%%%%%%%%%
%% Bibliography (alphabetical by key) %%
%%%%%%%%%%%%%%%%%%%%%%%%%%%%%%%%%%%%%%%%

\def\aa{A\&A} %% Astronomy and Astrophysics
\def\aar{A\&A~Rev.} %% Astronomy and Astrophysics Reviews
\def\aas{A\&AS} %% Astronomy and Astrophysics, Supplement
\def\actaa{Acta Astron.} %% Acta Astronomica
\def\aj{AJ} %% Astronomical Journal
\def\ao{Appl.~Opt.} %% Applied Optics
\def\apj{ApJ} %% Astrophysical Journal
\def\apjl{ApJ} %% Astrophysical Journal, Letters
\def\apjs{ApJS} %% Astrophysical Journal, Supplement
\def\apjsupp{ApJ Supp. Series} %% Astrophysical Journal, Supplement Series
\def\aplett{Astrophys.~Lett.} %% Astrophysics Letters
\def\apspr{Astrophys.~Space~Phys.~Res.} %% Astrophysics Space Physics Research
\def\apss{Ap\&SS} %% Astrophysics and Space Science
\def\araa{ARA\&A} %% Annual Review of Astron and Astrophys
\def\arxiv{arXiv} %% arXiv
\def\arxive{arXiv e-prints} %% arXiv, e-prints
\def\azh{AZh} %% Astronomicheskii Zhurnal
\def\baas{BAAS} %% Bulletin of the AAS
\def\bac{Bull. astr. Inst. Czechosl.} %% Bulletin of the Astronomical Institutes of Czechoslovakia
\def\bain{Bull.~Astron.~Inst.~Netherlands} %% Bulletin Astronomical Institute of the Netherlands
\def\caa{Chinese Astron. Astrophys.} %% Chinese Astronomy and Astrophysics
\def\cjaa{Chinese J. Astron. Astrophys.} %% Chinese Journal of Astronomy and Astrophysics
\def\cqg{CQGrav} %%Classical and Quantum Gravity
\def\ea{Exp. Astron.} %% Experimental Astronomy
\def\fcp{Fund.~Cosmic~Phys.} %% Fundamental Cosmic Physics
\def\gca{Geochim.~Cosmochim.~Acta} %% Geochimica Cosmochimica Acta
\def\gcn{GCN Circ. } %% GRB Coordinates Network
\def\grl{Geophys.~Res.~Lett.} %% Geophysics Research Letters
\def\iaucirc{IAU~Circ.} %% IAU Circulars
\def\icarus{Icarus} %% Icarus
\def\jcap{J. Cosmology Astropart. Phys.} %% Journal of Cosmology and Astroparticle Physics
\def\jcp{J.~Chem.~Phys.} %% Journal of Chemical Physics
\def\jgr{J.~Geophys.~Res.} %% Journal of Geophysics Research
\def\jqsrt{J.~Quant.~Spec.~Radiat.~Transf.} %% Journal of Quantitiative Spectroscopy and Radiative Trasfer
\def\jrasc{JRASC} %% Journal of the RAS of Canada
\def\memras{MmRAS} %% Memoirs of the RAS
\def\memsai{Mem.~Soc.~Astron.~Italiana} %% Mem. Societa Astronomica Italiana
\def\mnras{MNRAS} %% Monthly Notices of the RAS
\def\na{New A} %% New Astronomy
\def\nar{New A Rev.} %% New Astronomy Review
\def\nat{Nature} %% Nature
\def\nphysa{Nucl.~Phys.~A} %% Nuclear Physics A
\def\pasa{PASA} %% Publications of the Astron. Soc. of Australia
\def\pasj{PASJ} %% Publications of the ASJ
\def\pasp{PASP} %% Publications of the ASP
\def\physrep{Phys.~Rep.} %% Physics Reports
\def\physscr{Phys.~Scr} %% Physica Scripta
\def\planss{Planet.~Space~Sci.} %% Planetary Space Science
\def\pra{Phys.~Rev.~A} %% Physical Review A: General Physics
\def\prb{Phys.~Rev.~B} %% Physical Review B: Solid State
\def\prc{Phys.~Rev.~C} %% Physical Review C
\def\prd{Phys.~Rev.~D} %% Physical Review D
\def\pre{Phys.~Rev.~E} %% Physical Review E
\def\prx{Phys.~Rev.~X} %% Physical Review X
\def\prl{Phys.~Rev.~Lett.} %% Physical Review Letters
\def\procspie{Proc.~SPIE} %% Proceedings of the SPIE
\def\psci{PoS} %% Proceedings of Science
\def\qjras{QJRAS} %% Quarterly Journal of the RAS
\def\rmxaa{Rev. Mexicana Astron. Astrofis.} %% Revista Mexicana de Astronomia y Astrofisica
\def\skytel{S\&T} %% Sky and Telescope
\def\solphys{Sol.~Phys.} %% Solar Physics
\def\ssr{Space~Sci.~Rev.} %% Space Science Reviews
\def\sovast{Soviet~Ast.} %% Soviet Astronomy
\def\zap{ZAp} %% Zeitschrift fuer Astrophysik
\def\nucinstrummethodsa{Nucl.~Instrum.~Methods~A} %% Nuclear Instruments and Methods in Physics Research Section A: Accelerators, Spectrometers, Detectors and Associated Equipment

%%%%%%%%%%%%%%%%%%%%%%%%%%%%%%%%%%%%%%%%
%%           AASTeX & LaTeX           %%
%%%%%%%%%%%%%%%%%%%%%%%%%%%%%%%%%%%%%%%%

\newcommand\aastex{AAS\TeX}
\newcommand\latex{La\TeX}

%%%%%%%%%%%%%%%%%%%%%%%%%%%%%%%%%%%%%%%%
%%           Miscellaneous            %%
%%%%%%%%%%%%%%%%%%%%%%%%%%%%%%%%%%%%%%%%

\definecolor{burntorange}{rgb}{0.8, 0.33, 0.0}
\definecolor{amber}{rgb}{1.0, 0.75, 0.0}
\definecolor{ao(english)}{rgb}{0.0, 0.5, 0.0}
\definecolor{darkorchid}{rgb}{0.6, 0.2, 0.8}

\newcommand{\red}[1]{\textcolor{red}{\textit{#1}}}
\newcommand{\orange}[1]{\textcolor{burntorange}{\textit{#1}}}
\newcommand{\yellow}[1]{\textcolor{amber}{\textit{#1}}}
\newcommand{\green}[1]{\textcolor{ao(english)}{\textit{#1}}}
\newcommand{\blue}[1]{\textcolor{blue}{\textit{#1}}}
\newcommand{\purple}[1]{\textcolor{darkorchid}{\textit{#1}}}

\def\astar{A_\star}
\def\et3{\eta_3}
\def\th1{\theta_{-1}}
\def\r07{r_{0,7}}
\def\x05{x_{0.5}}
\def\muh{\hat{\mu}}
\def\cm{\hbox{~cm}}
\def\kpc{\hbox{~kpc}}
\def\Mpc{\hbox{~Mpc}}
\def\Gpc{\hbox{~Gpc}}
\def\s{\hbox{~s}}
\def\gev{\hbox{~GeV}}
\def\Jy{\hbox{~Jy}}
\def\Hz{\hbox{~Hz}}
\def\TeV{\hbox{~TeV}}
\def\GeV{\hbox{~GeV}}
\def\MeV{\hbox{~MeV}}
\def\kev{\hbox{~keV}}
\def\keV{\hbox{~keV}}
\def\eV{\hbox{~eV}}
\def\G{\hbox{~G}}
\def\erg{\hbox{~erg}}
\def\s{{\hbox{~s}}
\def\cm2{\hbox{~cm}^2}}
\def\para{\parallel}
\def\Fl{\mathcal{F}}
\def\Ep{E_{\rm peak}}
\def\Epk{E$_{\rm peak}$}
\def\Eiso{E$_{\rm iso}$}
\def\Liso{L$_{\rm iso}$}
\def\T90{T$_{\rm 90}$}
\def\tv{t_{\rm var}}
\def\Lt{L_{\hbox{erg s}^{-1}}}
\def\Lobs{{1.6 \times 10^{47}~ \hbox{erg s}^{-1}}}
\def\LobsOLD{{1.6 \times 10^{48}~ \hbox{erg s}^{-1}}} %check the numbers b/c this was  wrong
\def\r0{1.2 \times 10^{6}~ \hbox{cm}}
\def\es{\hbox{~erg s}^{-1}}
\def\fr#1#2{{{#1} \over {#2}}}

%%%%%%%%%%%%%%%%%%%%%%%%%%%%%%%%%%%%%%%%
%%                                    %%
%% Affiliations (alphabetical by key) %%
%%                                    %%
%%%%%%%%%%%%%%%%%%%%%%%%%%%%%%%%%%%%%%%%

\newcommand{\CSPAR}{\affiliation{Center for Space Plasma and Aeronomic Research, University of Alabama in Huntsville, Huntsville, AL 35899, USA}}
\newcommand{\GSFC}{\affiliation{NASA Goddard Space Flight Center, University of Maryland, Baltimore County, Greenbelt, MD 20771, USA}}
\newcommand{\Jacobs}{\affiliation{Jacobs Space Exploration Group, Huntsville, AL 35806, USA}}
\newcommand{\LSU}{\affiliation{Department of Physics and Astronomy, Louisiana State University, Baton Rouge, LA 70803 USA}}
\newcommand{\MPI}{\affiliation{Max-Planck-Institut f\"{u}r extraterrestrische Physik, Giessenbachstrasse 1, D-85748 Garching, Germany}}
\newcommand{\MSFC}{\affiliation{NASA Marshall Space Flight Center, Huntsville, AL 35812, USA}}
\newcommand{\MSFCAstro}{\affiliation{ST12 Astrophysics Branch, NASA Marshall Space Flight Center, Huntsville, AL 35812, USA}}
\newcommand{\NASA}{\altaffiliation{NASA Postdoctoral Fellow}}
\newcommand{\NYUAbuDhabi}{\affiliation{Center for Astro, Particle, and Planetary Physics, New York University Abu Dhabi}}
\newcommand{\PdiB}{\affiliation{Dipartimento Interateneo di Fisica dell'Università e Politecnico di Bari, Via E. Orabona 4, 70125, Bari, Italy}}
\newcommand{\SdiB}{\affiliation{Istituto Nazionale di Fisica Nucleare - Sezione di Bari, Via E. Orabona 4, 70125, Bari, Italy}}
\newcommand{\SPA}{\affiliation{Department of Space Science, University of Alabama in Huntsville, 320 Sparkman Drive, Huntsville, AL 35899, USA}}
\newcommand{\UAH}{\affiliation{University of Alabama in Huntsville, 320 Sparkman Drive, Huntsville, AL 35899, USA}}
\newcommand{\UCD}{\affiliation{School of Physics, University College Dublin, Belfield, Dublin 4, Ireland}}
\newcommand{\USRA}{\affiliation{Science and Technology Institute, Universities Space Research Association, Huntsville, AL 35805, USA}}
\newcommand{\UNAM}{\affiliation{Instituto de Astronom\' ia, Universidad Nacional Aut\'onoma de M\'exico, Circuito Exterior, C.U., A. Postal 70-264, 04510 Cd. de M\'exico, M\'exico.}}
%%%%%%%%%%%%%%%%%%%%%%%%%%%%%%%%%%%%%%%%
%%                                    %%
%%.  Contributing Authors (ranked)    %%
%%                                    %%
%%%%%%%%%%%%%%%%%%%%%%%%%%%%%%%%%%%%%%%%

%%%%%%%%%%%%%%%%%%%%%%%%%%%%%%%%%%%%%%%%
%%                                    %%
%%  Fermi-GBM Authors (alphabetical)  %%
%%                                    %%
%%%%%%%%%%%%%%%%%%%%%%%%%%%%%%%%%%%%%%%%

\author[0000-0002-2149-9846]{P.~Veres}
\SPA
\CSPAR

\author[0000-0002-0173-6453]{N.~Fraija}
\UNAM

\author[0000-0001-8058-9684]{S.~Lesage}
\SPA

\author[0000-0002-0587-7042]{A.~Goldstein}
\USRA
\author[0000-0003-2105-7711]{M.~S.~Briggs}
\SPA
\CSPAR
\author[0000-0001-7916-2923]{P.N.~Bhat}
\SPA
\CSPAR

%
%\author{friends}
%
%
%%
%\author[0000-0001-7916-2923]{P.N.~Bhat}
%\SPA
%\CSPAR
%%
%\author[0000-0001-9935-8106]{E. Bissaldi}
%\PdiB
%\SdiB
%%
%\author[0000-0003-2105-7711]{M.~S.~Briggs}
%\SPA
%\CSPAR
%%
%\author[0000-0002-2942-3379]{E.~Burns}
%\LSU
%%
%\author{W.~H.~Cleveland}
%\USRA
%%
%\author[0000-0003-3248-5447]{R.~Dunwoody}
%\UCD
%%
%\author{M.~M.~Giles}
%\Jacobs
%%
%\author{C.~Fletcher}
%\USRA
%%
%\author[0000-0002-0587-7042]{A.~Goldstein}
%\USRA
%%
%\author[0000-0001-9556-7576]{B.~Hristov}
%\UAH
%%
%\author[0000-0002-0468-6025]{C.~M.~Hui}
%\MSFC
%%
%\author{D.~Kocevski}
%\MSFC
%%
%\author[0000-0002-2531-3703]{B.~Mailyan}
%\NYUAbuDhabi
%%
%\author[0000-0002-0380-0041]{C.~Malacaria}
%\MSFC
%\USRA
%\NASA
%%
%\author[0000-0002-6269-0452]{S.~Poolakkil}
%\SPA
%%
%\author[0000-0003-1626-7335]{R.~Preece}
%\SPA
%%
%\author[0000-0002-7150-9061]{O.J.~Roberts}
%\USRA
%%
%\author[0000-0002-0221-5916]{A.~von Kienlin}
%\MPI
%%
%\author[0000-0002-8585-0084]{C. A. Wilson-Hodge}
%\MSFCAstro
%%
%
\newcommand{\grb}{GRB 221009A\xspace}

\newcommand{\SL}[1]{\textcolor{red}{[SL: #1]}}

\newcommand{\NF}[1]{\textcolor{green}{[NF: #1]}}

%% ABSTRACT 
\begin{abstract}
\noindent The IceCube neutrino observatory detects the diffuse astrophysical neutrino background with high significance, but the contribution of different classes of sources is not established. Because of their non-thermal spectrum, gamma-ray bursts (GRBs) are prime particle acceleration sites and one of the candidate classes for significant neutrino production. Exhaustive searches, based on stacking analysis of GRBs however could not establish the link between neutrinos and GRBs.
Gamma-ray burst GRB 221009A had the highest time integrated gamma-ray flux of any  detected GRB so far. The total fluence exceeds the sum of all Fermi Gamma-ray Burst Monitor (GBM) detected GRBs by a factor of two. 
Because it happened relatively nearby, it is one of the most favorable events for neutrino production from GRBs yet no neutrinos were detected. We calculate neutrino fluxes for this GRB in the TeV-PeV range using the most accurate, time-resolved spectral data covering the brightest intervals. We place limits on the physical parameters (Lorentz factor, baryon loading or emission radius) of the burst that are better by a factor of 2 compared to previous limits. The neutrino non-detection indicates a bulk Lorentz factor greater than $500$ and possibly even ${800} $, consistent with other observations.
\end{abstract}

%% INTRO 
\section{Introduction} \label{sec:intro}

%\noindent
%
Gamma-ray bursts (GRBs) have non-thermal photon spectra \citep[e.g.][]{Band+93,Poolakkil_2021} and for this reason are one of the most favorable candidates for non-electromagnetic signals, specifically very high energy (VHE) neutrinos. Gamma-rays are likely emitted by a population of accelerated electrons and protons will be accelerated alongside the electrons \citep[e.g.][]{2000ApJ...537..785D, 2000ApJ...541L...5M, 2000ApJ...537..255D, 2004A&A...418L...5D, 2004ApJ...604L..85A,
2009ApJ...705L.191A}. Protons accelerated up to ultra-high energies \citep[$> 10^{18}$ eV; ][]{1995PhRvL..75..386W, 1995ApJ...453..883V} can interact with the gamma-ray photons, suggesting GRBs are possible VHE neutrino sources \citep[e.g.][]{1997PhRvL..78.2292W, 2008PhRvD..78j1302M, 2009ApJ...691L..67W, 2013ApJ...772L...4G, 2013PhRvL.110l1101Z}.

%IceCube detected a diffuse, VHE neutrino flux at the level of $10^{-8} ~{\rm GeV}~ {\rm cm}^{-2} ~{\rm s}^{-1} ~{\rm sr}^{-1}$ \citep{ICE+13firstnuSCI}. 

%It has been continually reported and updated by the 
The IceCube Collaboration reported more than one hundred VHE  astrophysical neutrinos through the High-Energy Starting Events (HESE) catalog  \citep{2013PhRvL.111b1103A,2013Sci...342E...1I,2017arXiv171001191I, IceCube2020arXiv200109520I, IceCube_2021PhRvD.104b2002A}, and 276 events via the IceCube Event Catalog of Alert Tracks \citep{2023ApJS..269...25A}.  On the other hand, a VHE astrophysical muon–neutrino flux with events in the energy range of 15 TeV to 5 PeV during 9.5 yr of data taking was reported by IceCube Observatory \citep{2022ApJ...928...50A}.   The most suitable single power-law parametrization for the astrophysical events yields a normalization value of $1.44^{+0.25}_{-0.26}\times 10^{-18}\,{\rm GeV^{-1\,}cm^{-2}\,s^{-1}\,sr^{-1}}$ with a spectral index of $2.37\pm0.09$ \citep{2022ApJ...928...50A}.

Although several potential candidates have been suggested as possible neutrino sources  \citep[for a review see][]{2008PhR...458..173B, 2015RPPh...78l6901A}, there is currently no clear dominant population responsible for the diffuse neutrino flux. Currently the flaring blazar TXS 0506+056 \citep{Aartsen+18nublazarmma}, and the nearby Seyfert galaxy NGC1068 \citep{IceCube+22_NGC1068} stand out as significant point sources in the VHE neutrino sky, but individually these sources do not contribute a significant fraction of the diffuse flux.  The initial diffuse flux discovery and the subsequent point sources spurred an intense search for the class of objects that can explain the diffuse neutrino flux.

Blazars can contribute at most $\lesssim30\%$ of the diffuse neutrino background, based on stacking analyses of Fermi blazars \citep{Padovani2015MNRAS.452.1877P, Aartsen+17agnnu, Buson_2022ApJ...933L..43B}.  GRBs \citep{Aartsen+16ICGRB,Aartsen+17GRBAeff, 2014MNRAS.437.2187F}, tidal disruption events \citep{Stein+21nuTDE,Winter+21nuTDE} and star forming galaxies \citep{Tamborra+14nuStarburst} are some of the potential candidate classes, however it is not clear if any of these could have a dominant contribution. {Recently, Seyfert galaxies have been emerging as possible sources \citep{Murase+20nu_agn_hidden, Inoue+19nu_agn} after the increased neutrino signal from the direction of NGC 1068 \citep{Aartsen+20IceCube_pointsource}.}

Gamma-ray bursts have initially been considered as viable neutrino sources, however with increasingly more constraining observations, neutrinos from the prompt episode of GRBs as a population are more disfavored.  After performing an extensive examination of many years of data, the IceCube collaboration revealed a lack of any coincidences between GRBs and VHE neutrinos \citep{2022arXiv220511410A,2012Natur.484..351A, 2016ApJ...824..115A, 2015ApJ...805L...5A}.   Recently, \citet{IceCube+22GRBlimit} ruled out GRBs' contribution to the diffuse background at higher than $\sim10\%$ level for the early emission, considering Fermi bursts detected by the Gamma-ray Burst Monitor (GBM) instrument.  IceCube has regularly followed up transients:  GRB 080319B, the naked-eye burst \citep{Abbasi+09nunakedeye}; the highest
fluence GRB up to \grb, GRB~130427A \citep{Blaufuss+13_130427a_neutrinoGCN}; the first binary black hole merger detected in gravitational waves \citep[GW150914,][]{Icecube+16nuGW150914}, the first binary neutron star merger detected in both gamma-rays (GRB~170817A) and gravitational waves \citep[GW170817,][]{Albert+17nu170817a}. \citet{Gao+13nu130427a} explored the neutrino production of the bright GRB~130427A based on IceCube upper limits  and constrained the neutrino production parameters. We will use similar methods to rule out a portion of the parameter space.  
\cite{2018ApJ...859...70F} reported that the null neutrino result reported by IceCube observatory around GRB~080319B and GRB~130427A could be understood in relation to the degrees of magnetization \citep{2013PhRvL.110l1101Z}. \citet{2017ApJ...848...15F} considered the energy fraction given to accelerate cosmic rays and electrons in one of the brightest bursts GRB 160625B. After describing the multiwavelength observations, they found that the energy fraction converted into cosmic rays was drastically restricted by that given to accelerated electrons, thus explaining the lack of neutrino events in this burst.  

The observation of the nearby, extremely bright \grb together with non-detection of neutrinos offers unique ways to address the level of neutrino production in GRB models. While stacking analyses need to assume a distribution of Lorentz factors and estimates for the baryon loading parameter for each GRB, using a single GRB like \grb reduces this uncertainty as it relies on a single assumption instead of one for an entire population. \citet{Murase+22nuBOAT, Ai+23nuBOAT, Liu+23nuBOAT} investigated the implication on the neutrino models based on the time integrated gamma ray flux and the initial neutrino upper limit. Here we extend and improve upon these studies by using the time dependent gamma-ray flux of \grb from \citet{Lesage+23_221009a}, numerically integrated neutrino spectrum and the published IceCube upper limits for this GRB \citep{IceCube+23_grb221009a}.

In Section \ref{sec:obs} we describe the gamma-ray observations and the neutrino upper limits derived by the IceCube Collaboration. In Section \ref{sec:neutrino} we present the neutrino model calculations and the gamma-ray production. In Section \ref{sec:results} we present the results based on the constraints using the parameters of \grb, and in Section \ref{sec:discussion} we discuss these results and present our conclusions. We use cosmological parameters $\Omega_\Lambda=1-\Omega_m=0.714$, $H=67.4\pm 0.5 \km \s^{-1}\Mpc^{-1}$ \citep{2020A&A...641A...6P}, we use the notation $Q_{\rm x}=Q / 10^{\rm x}$ for quantity "Q" in {\it cgs} units and the usual notation for physical constants.

\section{Observations}\label{sec:obs}
\subsection{Gamma-rays}

\fermi-GBM detected \grb \citep{Veres+22_231009agcn32636, Lesage+22_221009agcn32642} and \citet {Lesage+23_221009a} reported the detailed gamma-ray properties. The prompt emission was further detected by a slew of instruments \citep[e.g.][]{gcn32637,gcn32641,An+23Insight_boat,Frederiks+23Konus_221009A,GRBAplha_221009A,Williams+23Swift_boat}, including a VHE energy detection by the Large High Altitude Air Shower Observatory \citep[LHAASO;][]{Huang+22LHASO_gcn32677, Cao+23LHAASO}.  The redshift of \grb is z=0.151 \citep{deugarte+2222109a_redshift_gcn32648} which corresponds to a luminosity distance of $D_L\approx 2.2\times10^{27}\cm$.  \grb consisted of multiple emission episodes described in \citet{Lesage+23_221009a}. The  fluence (time integrated gamma-ray flux in the 10-1000 keV range) of this powerful burst is dominated by the extremely bright episodes from  219 to 278 s  and  517 to 520 s after the trigger time (pulses 1, and 2 in Figure \ref{fig:specall}, respectively).  These intervals contain more than 90\% of the total flux.  In fact, the  fluence of \grb ($\sim 10^{-1} \erg\, \cm^{-2}$, \citet{Lesage+23_221009a}) exceeds the sum of the gamma-ray fluences from all GRBs detected by {\it Fermi}-GBM up to \grb by more than a factor of 2 \citep[see e.g.][for a discussion of the rate of similarly bright GRBs]{Burns+23boat}.

The extreme brightness during this phase exceeded the gamma-ray count rate where GBM operates nominally. The electronic pulses generated in the instrument by the gamma-rays piling up (the effect is called pulse pile-up or PPU) caused distortions in the spectrum.  \citet{Lesage+23_221009a} reconstructed the time-resolved spectra that were affected by the PPU 
%\SL
{using the analytical method developed by \cite{Chaplin2013} and verified by \cite{Bhat2014} to determine the spectral shape in each 1 s interval.}
%and determined the spectral shape in each 1 s interval. 
%\SL{should we add that these spectral results are not published?} 
To characterize the gamma-ray spectra, we use the parameters of the best fitting Band function \citep{Band+93} in each  interval, corrected for PPU. The Band function consists of two-, smoothly joined power law segments for the differential photon number spectrum: $dN_\gamma/dE_\gamma\propto E_\gamma^{\alpha}\exp{(-E_{\gamma}(\alpha+2)/E_{\rm peak})}$ if $E_{\gamma}<(\alpha+2)E_{\rm peak}/(\alpha+\beta)$ and $dN_\gamma/dE_\gamma\propto E_\gamma^{\beta}$ otherwise. Here $E_{\rm peak}$ is the peak energy in the energy-per-decade or $\nu F_\nu (=E_\gamma^2 dN_\gamma/dE_\gamma)$ representation, and $\alpha$ an $\beta$ the lower and higher spectral indexes, respectively. 
The reconstructed gamma-ray spectrum at different intervals is shown in Figure \ref{fig:specall}, plotted in the 10-10,000 keV range. 
{We present the PPU-corrected Band function spectral parameters for 1.024 s intervals in Table \ref{tab:specpar}. These values were not previously published in \citet{Lesage+23_221009a} because it is unclear if the Band function is the optimal spectral shape due to the complex nature of the PPU correction. 
We suggest contacting the authors of \citet{Lesage+23_221009a} for specific uses of these spectral parameters. Nonetheless the parameters in Table \ref{tab:specpar} represent an improved description of the spectrum. We note that the corrected fluence is consistent with the fluence derived by Konus-WIND \citep{Frederiks+23Konus_221009A}. 

% This is due to the complex nature of the PPU correction.% The spectral fit residuals could be due to spectral evolution of the data or uncorrected PPU effects. 
%We caution any reader who wishes to use these spectral fit results for scientific purposes and note it would be best to contact the authors of \citet{Lesage+23_221009a} before doing so.
}

The minimum variability timescale (MVT) is the timescale where coherent variations in the lightcurve can be identified. It is an important ingredient of e.g. internal shock models, relating the emission radius and the bulk Lorentz factor.
Because of the aforementioned data issues, the MVT during the brightest part of \grb is uncertain.  \citet{Lesage+23_221009a} determined the MVT outside the intervals affected by PPU (their Figure 2). The MVT varies between $10^{-2}-1 \s$ throughout the GRB, and generally is lower for brighter intervals. This means that even though it is not directly measured, $10^{-2}\s$ is a reasonable assumption. For our calculations we will assume  $\Delta t_{\rm var}=10^{-2}\,\s$ but present results using $\Delta t_{\rm var} =10^{-1}\, \s$ as well.

\subsection{Neutrino upper limits}
The IceCube Collaboration reported initial non-detection limits for \grb in \citet{IceCube221009a_gcn32665} and more detailed analysis in \citep{IceCube+23_grb221009a}. Neutrino spectrum with different power law slopes result in different limits  (see the IceCube  sensitivity in Figure \ref{fig:specall}). 

They found that at the time of \grb neutrino fluence could be constrained at the $(3-4)\times 10^{-2} \GeV \cm^{-2}$ level in the TeV-PeV energy range. In this paper, we focus on this neutrino energy range and do not consider limits at lower energies \citep[see e.g.][]{Murase+22nuBOAT,IceCube+23_grb221009a}.

\section{Neutrino model constraints} \label{sec:neutrino}
%\section{Neutrino flux model}

The most widely used scenario for explaining the properties of GRB prompt emission is the internal shock model \citep{Rees+94is}. Instabilities in the highly relativistic jet lead to collisions of shells inside the jet that drive shocks, accelerate particles and enhance the magnetic fields. Shock accelerated electrons in magnetic fields emit synchrotron radiation which is responsible for the observed prompt gamma-rays \citep{1995ApJ...455L.143S,1996ApJ...473..204S}. Protons are similarly shock accelerated and for a non-negligible time share the same volume as the gamma-rays. The accelerated protons and gamma-ray photons will interact and produce VHE neutrinos \citep[e.g.][]{1997PhRvL..78.2292W, 2008PhRvD..78j1302M, 2009ApJ...691L..67W, 2013ApJ...772L...4G, 2013PhRvL.110l1101Z}. 

Because of the relation between the photon and proton energies in the main channel of the $p\gamma$ process (the product of the proton and photon energies is approximately constant), the low-energy neutrino spectral slope will `inherit' the high-energy  photon index of the Band function ($dN_\nu/dE_\nu \propto E_\nu^{s-\beta-1}$) and the high-energy neutrino slope will be defined by the low-energy photon spectral index ($dN_\nu/dE_\nu \propto E_\nu^{s-\alpha-1}$). At higher neutrino energies the impact of pion (or muon) cooling will result in an additional spectral break  ($dN_\nu/dE_\nu \propto E_\nu^{s-\alpha-3}$, e.g. \citet{Zhang+13nu}). The proton number distribution is characterized by the index $s$, $dN_p/dE_p\propto E_p^{s}$.

\subsection{Detailed derivation }

The $p\gamma$ process 
results in three neutrinos through pion and muon decay. The energy of each neutrino in the frame comoving with the jet ($\ve_\nu$) is about 1/20 of the parent proton energy ($E_p$) \citep[e.g., see][]{1997PhRvL..78.2292W}. 
Here, we describe the photon and the proton energy distribution and calculate the relevant timescales for the neutrino production from their interactions.

\subsubsection{Photons} 
We calculate the photon spectral density, $dn(\ve_\gamma)/d\ve_\gamma$  from the observed spectra and the associated flux. 
The photon number density comoving with the jet will have the same shape as the observed spectrum, shifted in energy by $(1+z)/\Gamma$ \citep{1997PhRvL..78.2292W}:
\begin{equation}
    \frac{dn(\ve_\gamma)}{d\ve_\gamma}\propto \left\{
    \begin{array}{ll}
    {\ve_\gamma}^{\alpha} \exp \left[ -\frac{(\alpha + 2) \ve_\gamma}{\ve_{\gamma,peak}}  \right], & {\ve_\gamma} \geq  \ve_{\gamma,c}  \\
                \ve_\gamma^{\beta}, & {\ve_\gamma} < {\ve_{\gamma,c}}\,,
               \end{array}
             \right.
\end{equation}
where the comoving cutoff energy ($\ve_{\gamma,c}$) is related to the peak energy by
$\ve_{\gamma,c}={(\alpha - \beta) \ve_{\rm \gamma,peak}}/{(\alpha +2)} $. 

It is normalized so that $U_\gamma=\int \ve_\gamma dn(\ve_\gamma)$ in a given time interval equals to $L_\gamma/4\pi R^2 c \Gamma^2$, where $\Gamma$ is the bulk Lorentz factor and $c$ is the speed of light. The gamma-ray luminosity ($L_\gamma$) is calculated for each 1 ~s interval by performing k-correction to the canonical 1-10,000 keV range. In the internal shock model \citep{Rees+94is}, the radius where the gamma-rays are produced is $R = {2}\Gamma^2 c \Delta t_{\rm var} /(1+z)$, where $\Delta\, t_{\rm var}$ is the observed variability of the prompt gamma-ray lightcurve and $z$ is the redshift.

\subsubsection{Protons} 
The shocked proton distribution $(dN_p/dE_p)$ will extend from a minimum proton energy of $E_{p, {\min}}\approx\Gamma m_p c^2$ to a maximum energy $E_{p, \max}$ \cite[e.g., see][]{2008PhRvD..78j1302M, 2009ApJ...691L..67W, 2014MNRAS.437.2187F}. $E_{p, \max}$ is defined by the competition between shock acceleration and adiabatic timescales, proton synchrotron or $p\gamma$ cooling. We assume the power law index is $s = -2$. {The energy in shocked protons is related to the observed gamma-ray energy through the baryon loading parameter $\xi_p = E_{{\rm p, iso}}/E_{\gamma,{\rm iso}}$, where $E_{\gamma,{\rm iso}}$ is the isotropic-equivalent gamma-ray energy, calculated separately for each 1 s time bin. We assume $\xi_p$ is the same over all intervals. The total energy for \grb is $E_{\gamma,{\rm iso}}^{\rm total}= 10^{55}$ erg \citep{Lesage+23_221009a}.}

\subsubsection{Neutrino production}
The accelerated protons will undergo $p\gamma$ interaction with the photons of the prompt emission and will create charged ($\pi^+$) and neutral  pions ($\pi^0$):
$p+\gamma\rightarrow n\,+\, \pi^{+}$ and $ p\,+\,\pi^0$, respectively. The charged pions decay into muons ($\mu^{+}$) and muon neutrinos ($\nu_{\mu}, \bar{\nu}_\mu$): $\pi^{+}\rightarrow\mu^{+}+\nu_\mu (\rm{or}~ \bar{\nu}_\mu)$, and  the neutral pions decay into two photons \cite[e.g., see][]{2008PhRvD..78j1302M, 2009ApJ...691L..67W, 2014MNRAS.437.2187F}.

Protons undergoing $p\gamma$ process lose their energy at a rate of
\begin{equation}
\label{eq:tpgamma}
    t_{p\gamma}^{-1} (\gamma_p) = \frac{c}{2\gamma_p^2}\int\limits_{\bar{\ve}_{\rm thr}}^{\infty}
    d \bar{\ve}_\gamma~ \sigma_{p\gamma}(\bar{\ve}_\gamma)~  \xi(\bar{\ve}_\gamma)
    \bar{\ve}_\gamma
    \int\limits_{\bar{\ve}_\gamma/2\gamma_p}^{\infty}
    d\ve_\gamma \ve_\gamma^{-2} \frac{dn(\ve_\gamma)}{d\ve_\gamma}\,,
\end{equation}

\begin{figure}
    \centering
    \includegraphics[width=\columnwidth]{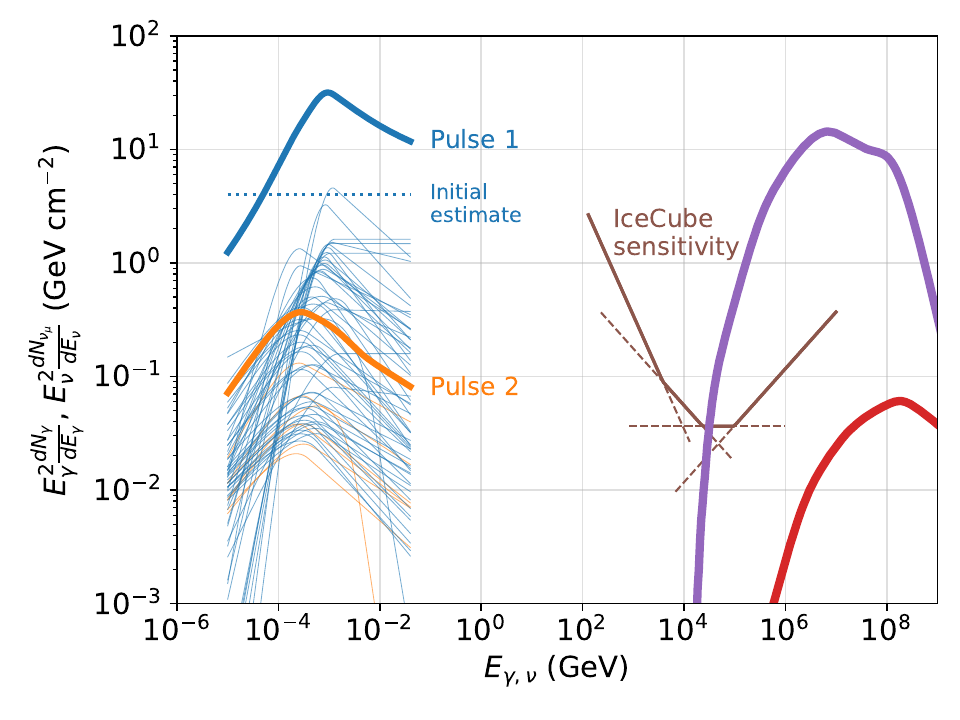}
    \caption{In the MeV range we show the gamma-ray spectrum. The thin lines represent spectra for 1 s time bins, the thick lines show the sum of each bin in the two most significant pulses. The horizontal blue dotted line shows the flat spectrum used by \citet{IceCube+23_grb221009a}. We show the neutrino spectra in purple and red  for the two example cases in Figure \ref{fig:IScontour} for ruled out and allowed parameter sets respectively. The brown curve shows the combined power law limits (90\%) for different power law neutrino spectra reported by \citet{IceCube+23_grb221009a}.}
    \label{fig:specall}
\end{figure}

\begin{figure}
    \centering
    \includegraphics[width=\columnwidth]{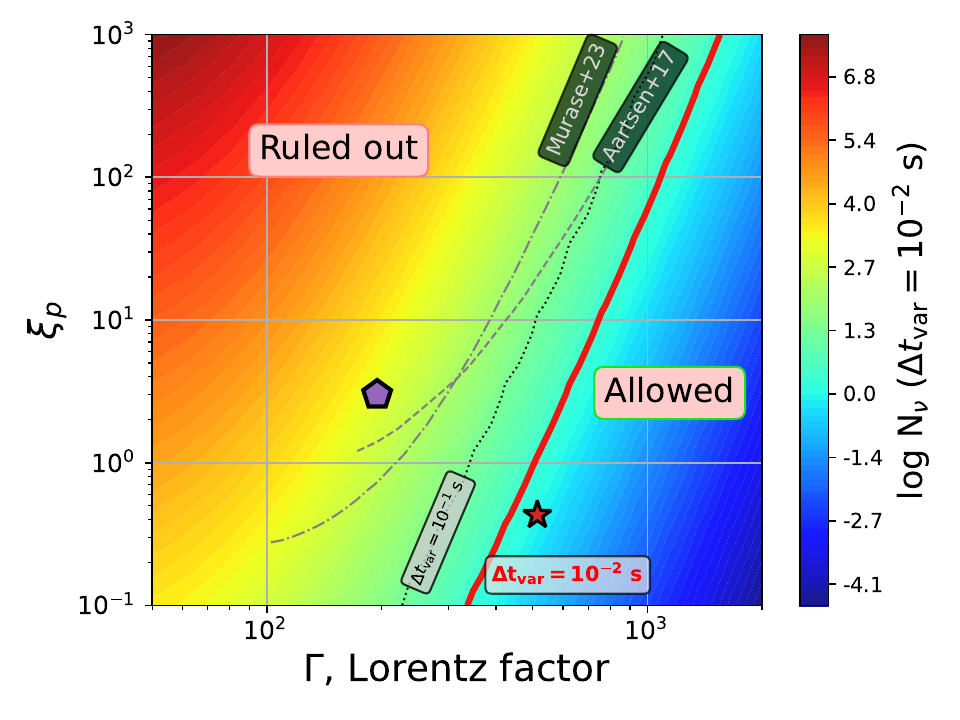}
    \caption{Constraints based on the internal shock model. The number of neutrinos are calculated for variability timescale of $\Delta t_{\rm var}=10^{-2}$~s. The red contour shows the 90\% confidence (2.44 neutrinos), the dotted lines indicate the 90\%  confidence level for $\Delta t_{\rm var}=10^{-1}$~s. The dash-dotted line shows the calculation of \citet{Murase+22nuBOAT}, while the stacking analysis of \citet{Aartsen+17GRBAeff} is shown with a dashed line. The two example cases are indicated by the purple pentagon and the red star, and the neutrino spectrum is shown in Figure \ref{fig:specall} with the corresponding colors.  }
    \label{fig:IScontour}
\end{figure}

where $\gamma_p=E_p/({\Gamma} m_p c^2)$ is the proton random Lorentz factor, $ \bar{\ve}_\gamma$ is the photon energy in the frame of the protons,  $\sigma_{p\gamma}$ is the $p\gamma$ cross section and $\xi$ is the inelasticity. 
{It is known that the multi-pion channel increases the neutrino flux by a factor of 2-3 \citep{He+12neutrino,Murase+22nuBOAT,Kimura+22grbnu}. In our treatment we use the exact cross section, composed of multiple neutrino production channels \citep{pdg+08pgamma_cross}. However for simplicity, we approximate the inelasticity by $\xi=0.2$ for $\bar{\ve}_\gamma<980$~MeV (mainly $\Delta$ resonance and direct production) and $\xi=0.6$  for $\bar{\ve}_\gamma>980$~ MeV (multi-pion channel) \citep{atoyander01-nuagn,Hummer+10pgamma}. }

The expansion of the jet on a dynamic timescale $t_{\rm dyn}\simeq R/\Gamma c$, will reduce the efficiency of the p$\gamma$ process. This can be described as a multiplicative suppression factor $f_{p,\gamma} =1-\exp(- t_{\rm dyn}/t_{p\gamma}$). 
We  numerically integrate Equation \ref{eq:tpgamma} with the appropriate photon and proton distribution to obtain $f_{p\gamma}$. 

Depending on the strength of the magnetic field in the region where the p$\gamma$ interaction occurs, the muons or pions can cool  significantly through synchrotron radiation before producing neutrinos. This will also result in a softening of the high-energy neutrino spectrum, encapsulated by additional suppression factors $f_{\pi},  f_{\mu}$: 
\begin{equation}
f_{\pi,\mu}(E_\nu)=1-\exp{-(t^{\rm cool}_{\pi,\mu}/t^{\rm decay}_{\pi,\mu})}\,,
\end{equation}
where the cooling timescale in magnetic field $B$ will be 
$t^{\rm decay}_{\pi}= 2.6\times 10^{-8} \gamma_\pi$, where $\gamma_\pi$ is the random Lorentz factor of the pions, which inherit 20\% of the proton's energy.  The decay timescale  $t^{\rm decay}_{\mu}=2.2\times 10^{-6} \gamma_\mu \s $ where $\gamma_\mu$ is the random Lorentz factor of the muons, that inherit 15\% of the proton's energy. The synchrotron cooling time in the comoving frame will be $t^{cool}_{\pi,\mu}={6\pi m_{\pi,\mu}^3 c}/{\gamma_{\pi,\mu} m_e \sigma_T B^2}$, where $m_\pi$, $m_\mu$ and $m_e$ correspond the pion, muon and electron masses, respectively, and $\sigma_T$ the Thomson cross section.  The magnetic field energy density is a fraction $\epsilon_B$ of the gamma-ray energy density; i.e. $B^2/8\pi=\epsilon_B L_\gamma/4\pi c\Gamma^2 R^2$.  We consider the value of $\epsilon_B=10^{-2}$, and note that it only slightly affects the calculated number of neutrinos.  %

With the suppression factors above, after taking into account neutrino oscillations, the {muon}  neutrino flux becomes
\begin{equation}
\label{eq:nu}
   E_{\nu_\mu}^2 \frac{dN_{\nu_\mu}}{dE_{\nu_\mu}} = \frac{\mathcal{R}}{4} f_{p\gamma} (0.4\, f_{\mu}+0.6 f_{\mu}  f_{\pi}) E_p^2 \frac{dN_p}{dE_p}.
\end{equation}
{Here $\mathcal{R}=1/2$ represents the fraction of charged pions to the number of total pions. For simplicity, we approximate $\mathcal{R}=1/2$ for the $\Delta$ resonance and $\mathcal{R}=2/3$ for the multi-pion channels  \citep{Hummer+10pgamma, He+12neutrino}. For more accurate treatment, numerical treatment is required \citep[e.g.][]{Mucke+99pgamma,murase+nag06prd,Baerwald+11grbnu}, we prefer the above approximation for ease of use and speed of calculation.}

\subsection{Number of expected neutrinos}
\label{sec:number_of_neutrinos}
We calculate the number of expected neutrinos, by convolving the neutrino spectrum with the IceCube effective area reported in \citet{IceCube+21effectivearea} using a declination of 19.8 degrees.  We compare the above effective area with the one reported specific for \grb \citep[see][]{IceCube+23_grb221009a} and found that the specific effective area is marginally, but systematically larger. They match within $\sim 10\%$ in the 10 TeV -10 PeV range. By using the generic effective area, we will slightly underestimate the number of neutrinos.

The reported upper limits for neutrino searches are typically at 90\% confidence. This corresponds to 2.3 neutrinos in Poisson statistics and 2.44 neutrinos in the Feldman-Cousins \citep{FeldmanCousins98statistics}  statistics. For comparison with previous studies \citep[e.g.][]{Murase+22nuBOAT} we use 2.44 neutrinos as 90\% confidence detection.

\section{Results}
\label{sec:results}
We calculate the neutrino spectrum (Equation \ref{eq:nu}) for a grid of parameters, by integrating over the time-dependent gamma-ray spectrum. We then convolve the neutrino spectrum with the effective area of IceCube to obtain the number of neutrinos ($N_\nu$). We rule out regions of parameter space based on the non-detection of neutrinos by IceCube given its sensitivity: parameters that result in a neutrino spectrum that is above the sensitivity curve, or equivalently result in more than 2.44 neutrinos can be ruled out, because there was no detection. Conversely, any parameter set that yields a neutrino spectrum below the IceCube sensitivity or fewer than 2.44 counts is still allowed.

\subsection{Internal shock constraints}
In the internal shock scenario, we fix the variability timescale, MVT and scan the bulk Lorentz factor ($\Gamma$), baryon loading ($\xi_p$) parameter space.   Given the used value of bulk Lorentz factor and variability timescale,  the dissipation radius becomes $R= 2.4\times 10^{14} ~\Gamma_{2.8}^2 \Delta t_{{\rm var},-2}\cm$.
We explore the range of $10^{-1}<\xi_p<10^3$ and $50\leq\Gamma\leq 2\times 10^3$ (Figure \ref{fig:IScontour}).
Assuming the MVT is $10^{-2}\s$, the 90\% confidence line between the allowed and the ruled out part of the $\Gamma-\xi_p$ parameter space is approximately $\xi_p\approx {150} \times(\Gamma/1000)^{6}$. This line runs parallel with- and improves upon previous studies for this GRB \citep{Murase+22nuBOAT} or stacked GRBs \citep{Aartsen+17GRBAeff}. Assuming a larger MVT,  e.g. $\Delta t_{\rm var}=10^{-1}\s$, the limit still improves over previous analyses (dotted line in Figure \ref{fig:IScontour}).

Furthermore, we illustrate two specific cases with parameters chosen from the allowed and the ruled out region.
The purple pentagon in Figure \ref{fig:IScontour} ($\Gamma \approx 200$, $\xi_p\approx3$) is rejected, because it produces a large number of neutrinos ($N\approx 1000$). Indeed the neutrino spectrum lies well above the published IceCube sensitivity (brown curve in Figure \ref{fig:specall}).
The red star in Figure \ref{fig:IScontour}  ($\Gamma \approx 500$, $\xi_p \approx0.4$) is in the allowed region because the number of neutrinos is below the 90\% threshold. Conversely, for this set of parameters the neutrino spectrum in Figure \ref{fig:specall} lies below the IceCube sensitivity.

\subsection{Model independent constraints}

\begin{figure}
    \centering
    \includegraphics[width=\columnwidth]{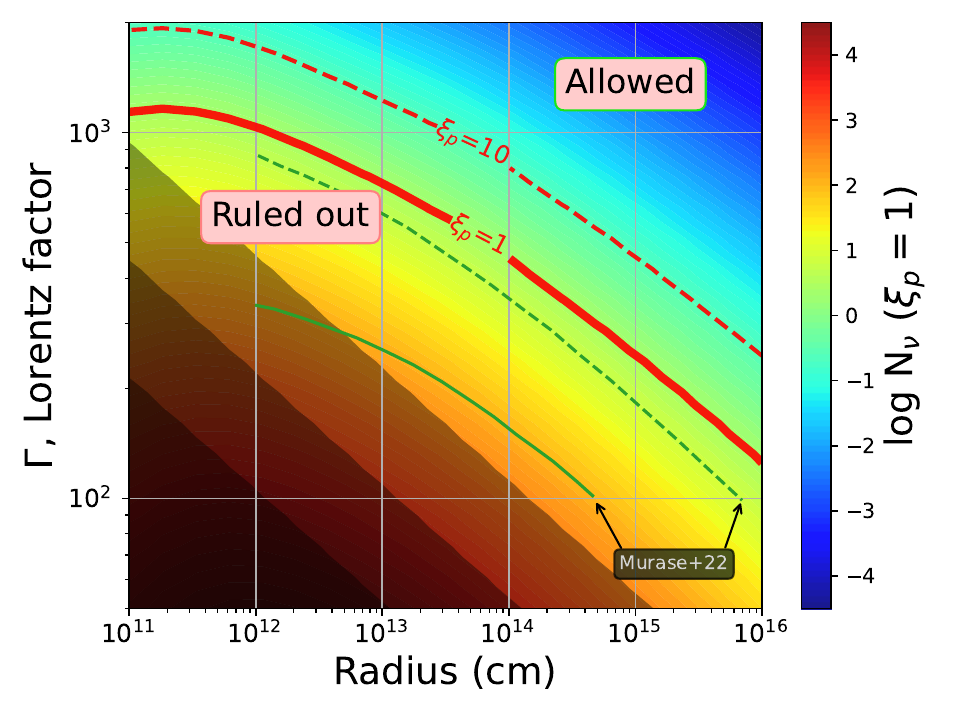}
    \caption{Constraints on the GRB parameter space without assuming a relationship between the Lorentz factor and radius. The red lines indicate the 90\% confidence region for $\xi_p=1$ (solid) and $\xi_p=10$ (dashed). 
    Green lines are from \citet{Murase+22nuBOAT} (solid: $\xi_p=1$ and dashed: $\xi_p=10$).  Shaded regions represent the photosphere for { proton} luminosities $10^{53}, 10^{52}$ and $10^{51} \erg \s^{-1}$ (light to darker shades).}
    \label{fig:general}
\end{figure}

In the model independent case, we assume a baryon loading parameter and scanned the bulk Lorentz factor ($\Gamma$) and emission radius ($R$) plane. 
We calculate the neutrino flux without any assumption about the relation between the radius and the bulk Lorentz factor.  In Figure \ref{fig:general} we show the allowed and ruled out parameter space for $\xi_p=1$ and  $\xi_p=10$, and compare it to \citet{Murase+22nuBOAT}.

The radius of the jet photosphere can be used to constrain the allowed parameter space. Denoting the ratio of electrons and positrons to protons by $\zeta_e$, the proton luminosity by $L_p$, the photospheric radius is   $R_{\rm ph}=\zeta_e L_p \sigma_T/4\pi m_p c^3 \Gamma^3=4.7\times 10^{11} L_{p,53} \zeta_{e,0} \Gamma_{2.8}^{-3} \cm$.  The borders of the 3 shaded regions in Figure \ref{fig:general} from lighter to darker are defined by $L_p=\{10^{53},\,10^{52} ~{\rm and}~ 10^{51}\}$ erg s$^{-1}$. Based on Figure \ref{fig:general}, in order to have no neutrino detection  we either need a large bulk Lorentz factor ($\Gamma \gtrsim 1000$) or a large dissipation radius.

\section{Discussion and Conclusion}
\label{sec:discussion}
We calculated the neutrino flux from the exceptionally bright \grb using the best available, time resolved description of the prompt gamma-ray spectrum. We found that large swaths of the parameter space can be ruled out, improving previous limits based on neutrino non-detection.

{The pulse-pileup correction for in the Fermi-GBM data of \grb will be affected by systematic uncertainties. However, based on the fact that the total fluence agrees with Konus-Wind \cite[$1.5\times 10^{55}\erg$;][]{Frederiks+23Konus_221009A} and it is consistent with Insight-HXMT \cite[$1.2\times 10^{55}\erg$;][]{An+23Insight_boat}, we assess the derived value is close to the actual energetics.} 
%%%%%%%%%%%%%%%%%%%%%%%%%%

Based  on the range in the reported values for the energetics, we estimated, a systematic uncertainty about 15\%. This uncertainty is compensated by the fact that we considered only the brightest part of \grb, containing about 90\% of the fluence, and by taking the more conservative IceCube effective area, which is $\sim 10\%$ below the one reported in \cite{IceCube+23_grb221009a}.

As a consistency check we compared the two approaches for claiming detection in IceCube: first, the number of neutrino events obtained by convolving the neutrino spectrum with the effective area needs to be above the 90\% confidence level of Poisson or Feldman Cousins statistic (Figures \ref{fig:IScontour} and \ref{fig:general}). At the same time, the neutrino spectral energy distribution needs to be above the 90\% differential sensitivity curve from e.g. \citet{IceCube+23_grb221009a}, as shown in  Figure \ref{fig:specall}. We found that the two approaches were consistent and illustrate it with two cases.

Our exclusion contour line runs parallel to that of \citet{Murase+22nuBOAT} that addresses \grb and \citet{Aartsen+17GRBAeff} that uses a stacking analysis of a larger sample of GRBs. Both the $\Delta t_{\rm var}=10^{-2}\s$ line and the more conservative  $\Delta t_{\rm var}=10^{-1}\s$ 90 \% contour lines visibly improve on the limits of previous works. This can be understood in part from the larger gamma-ray total energy used here, {$10^{55}$~erg  versus $10^{54.5}$~erg in } 
\citet{Murase+22nuBOAT} and in part due to the detailed spectral reconstruction.
{We note that the time resolved spectra (Table \ref{tab:specpar}) have systematically lower peak energies than 1.2 MeV and the high energy slope, beta, is systematically harder in the time resolved phase compared to the value of -2.66 employed by \citet{Murase+22nuBOAT}.}

If in our calculations we use $E_{\rm iso} =10^{54.5} \erg$, and a single time bin, with spectral parameters $E_{\rm peak}=1.2\MeV$, $\alpha=-0.89$ and $\beta=-2.66$ the limit corresponding to  $\Delta t_{\rm var}=10^{-2}\s$ will be consistent with that of \citet{Murase+22nuBOAT} (Figure \ref{fig:IScontour}). Small differences might be explained by the choice of the gamma-ray spectra described by two power laws with sharp break in \citet{Murase+22nuBOAT} compared  the smooth break of the Band function in this work. Furthermore, the fluence of \grb exceeds (or is comparable) the entire sample of GBM GRBs and it is closer than the median GRB (with measured redshift). These are the relevant drivers of neutrino flux in a typical stacking analysis where the gamma-ray contribution is taken from GRB catalogs.

We highlight that if the dissipation radius is kept fixed, knowing the detailed spectrum improved upon the exclusion limits of previous studies by a factor of ${\sim 2}$ on the value of the Lorentz factor. The improvement is approximately a factor of {30} in the allowed dissipation radii if the bulk Lorentz factor is kept constant (Figure \ref{fig:general}).

The most straightforward conclusion based on the non-detection of neutrinos from \grb is the requirement of a large value of the bulk Lorentz factor. The internal shock assumption with baryon loading close to unity yields $\Gamma\gtrsim 500$ (Figure \ref{fig:IScontour}), in line with \citet{Cao+23LHAASO}, based on the peak of the TeV afterglow emission. The model independent study (Figure \ref{fig:general}), again with baryon loading $\xi_p\approx 1$, and assuming a dissipation radius of $R=10^{14}\cm$, leads to similar constraints, $\Gamma\gtrsim 500$.  It is worth noting that if $R\lesssim10^{13}\cm$ we get $\Gamma\gtrsim {800}$.

Given that \grb is the brightest GRB of all time, it happened relatively nearby and did not yield positive neutrino identification, does not bode well for future coincident GRB-neutrino detection prospects. Here we concluded that the non-detection points to a relatively large Lorentz factor, close to 1000 or a low baryon load. The possibility of detection increases if future nearby GRBs have low bulk Lorentz factors which place the neutrino spectral peak at lower energies, closer to the sensitive region of IceCube. Lower Lorentz factor is synonymous with higher baryon load (dirtier fireball), which in turn may result in more favorable neutrino production. In other words, high-energy neutrinos might still be detected from future nearby GRBs with a low value of bulk Lorentz factor.

%%% ACKNOWLEDHEMENTS 

We thank Kohta Murase, Joshua Wood and P\'eter M\'esz\'aros for discussions on this topic {and the anonymous referee for their thoughtful and constructive  comments}.   P.~V., S.~L.,  M.~S.~B. and P.~N.~B. gratefully acknowledge NASA funding from cooperative agreement 80MSFC22M0004.  A.~G. gratefully acknowledges NASA funding through cooperative agreement 80NSSC24M0035. N.~F. is grateful to UNAM-DGAPA-PAPIIT for the funding provided by grant IN119123.

{\it Software}: numpy \citep{vanderWalt+11numpy}, matplotlib \citep{Hunter07matplotlib}, astropy \citep{astropy+18}, scipy \citep{Scipy+20}, GBM data tools \citep{GbmDataTools}.

%%% REFERENCES
\bibliographystyle{aasjournal}
%\bibliography{references, nu, grb}

%\usepackage{appendix}
\appendix

\startlongtable
\begin{deluxetable}{ccccc}
\label{tab:specpar}
\tabletypesize{\footnotesize}
\tablecolumns{5}
%\tablewidth{0pt}
\tablecaption{Parameters of spectral fits obtained after performing pulse pile-up corrections in the brightest regions of \grb. $A$ is the normalization at 100 keV of the photon spectrum, $dN_\gamma/dE_\gamma$. }
\tablehead{
\colhead{T$_{\rm start}$-T$_{\rm start}$} & \colhead{A} & \colhead{E$_{\rm peak}$  }& \colhead{$\alpha$} & \colhead{$\beta$} \\
\colhead{ s} & \colhead{ph/cm$^{2}$/s$^{}$/keV$^{}$} & \colhead{keV}& \colhead{} & \colhead{} \\}
\startdata
218.50 - 219.52 &  2.084 $^{+0.020}_{-0.020}$ 	& 1208.5 $^{+30.0}_{-30.5}$ 	& -1.095 $^{+0.009}_{-0.009}$ 	& -2.052 $^{+0.029}_{-0.028}$ \\
219.52 - 220.55 &  3.044 $^{+0.027}_{-0.027}$ 	& 1607.5 $^{+37.4}_{-37.6}$ 	& -1.070 $^{+0.007}_{-0.007}$ 	& -2.000 $^{+0.009}_{-0.000}$ \\
220.55 - 221.57 &  7.803 $^{+0.075}_{-0.076}$ 	& 1233.4 $^{+19.9}_{-20.0}$ 	& -0.941 $^{+0.007}_{-0.007}$ 	& -2.010 $^{+0.015}_{-0.010}$ \\
221.57 - 222.60 &  14.261 $^{+0.076}_{-0.313}$ 	& 874.3 $^{+12.7}_{-12.1}$ 	& -1.023 $^{+0.001}_{-0.054}$ 	& -2.188 $^{+0.023}_{-0.011}$ \\
222.60 - 223.62 &  13.543 $^{+0.117}_{-0.175}$ 	& 1294.2 $^{+1.4}_{-1.4}$ 	& -1.183 $^{+0.007}_{-0.007}$ 	& -2.663 $^{+0.032}_{-0.047}$ \\
223.62 - 224.64 &  28.497 $^{+0.309}_{-0.291}$ 	& 453.8 $^{+4.2}_{-4.4}$ 	& -0.898 $^{+0.014}_{-0.014}$ 	& -2.367 $^{+0.014}_{-0.015}$ \\
224.64 - 225.67 &  29.237 $^{+0.341}_{-0.290}$ 	& 628.5 $^{+6.2}_{-5.4}$ 	& -0.748 $^{+0.012}_{-0.011}$ 	& -2.284 $^{+0.012}_{-0.012}$ \\
225.67 - 226.69 &  13.387 $^{+0.146}_{-0.182}$ 	& 1068.9 $^{+7.8}_{-11.2}$ 	& -0.387 $^{+0.008}_{-0.009}$ 	& -2.000 $^{+0.003}_{-0.000}$ \\
226.69 - 227.72 &  4.896 $^{+0.049}_{-0.059}$ 	& 1157.2 $^{+6.3}_{-6.4}$ 	& 0.371 $^{+0.011}_{-0.007}$ 	& -2.110 $^{+0.008}_{-0.011}$ \\
227.72 - 228.74 &  4.397 $^{+0.048}_{-0.048}$ 	& 1114.8 $^{+6.9}_{-5.9}$ 	& 0.495 $^{+0.009}_{-0.009}$ 	& -2.001 $^{+0.046}_{-0.001}$ \\
228.74 - 229.77 &  6.133 $^{+0.033}_{-0.035}$ 	& 846.4 $^{+1.5}_{-2.3}$ 	& 1.496 $^{+0.006}_{-0.006}$ 	& -2.500 $^{+0.013}_{-0.013}$ \\
229.77 - 230.79 &  4.670 $^{+0.056}_{-0.048}$ 	& 1293.3 $^{+8.8}_{-7.3}$ 	& 0.273 $^{+0.009}_{-0.007}$ 	& -2.000 $^{+0.008}_{-0.000}$ \\
230.79 - 231.81 &  2.164 $^{+0.008}_{-0.012}$ 	& 1180.3 $^{+2.5}_{-1.8}$ 	& 1.647 $^{+0.004}_{-0.003}$ 	& -2.404 $^{+0.036}_{-0.003}$ \\
231.81 - 232.84 &  4.289 $^{+0.053}_{-0.041}$ 	& 966.4 $^{+4.4}_{-4.4}$ 	& 0.780 $^{+0.008}_{-0.011}$ 	& -2.339 $^{+0.012}_{-0.013}$ \\
232.84 - 233.86 &  5.787 $^{+0.011}_{-0.373}$ 	& 946.8 $^{+6.9}_{-3.5}$ 	& 0.532 $^{+0.010}_{-0.011}$ 	& -2.512 $^{+0.021}_{-0.013}$ \\
233.86 - 234.89 &  11.142 $^{+0.143}_{-0.127}$ 	& 905.2 $^{+5.5}_{-5.6}$ 	& -0.001 $^{+0.014}_{-0.008}$ 	& -2.599 $^{+0.016}_{-0.019}$ \\
234.89 - 235.91 &  14.159 $^{+0.158}_{-0.191}$ 	& 796.2 $^{+5.2}_{-4.9}$ 	& -0.068 $^{+0.022}_{-0.006}$ 	& -2.625 $^{+0.018}_{-0.018}$ \\
235.91 - 236.93 &  40.527 $^{+0.477}_{-0.444}$ 	& 457.2 $^{+3.9}_{-3.5}$ 	& -0.843 $^{+0.017}_{-0.016}$ 	& -2.657 $^{+0.018}_{-0.019}$ \\
236.93 - 237.96 &  30.255 $^{+0.031}_{-3.242}$ 	& 487.2 $^{+7.4}_{-6.2}$ 	& -1.380 $^{+0.011}_{-0.016}$ 	& -2.644 $^{+0.025}_{-0.028}$ \\
237.96 - 238.98 &  32.055 $^{+0.339}_{-0.327}$ 	& 269.5 $^{+2.8}_{-2.9}$ 	& -1.230 $^{+0.020}_{-0.020}$ 	& -2.484 $^{+0.017}_{-0.016}$ \\
238.98 - 240.01 &  39.802 $^{+0.374}_{-0.510}$ 	& 281.3 $^{+3.1}_{-3.5}$ 	& -1.351 $^{+0.019}_{-0.022}$ 	& -2.550 $^{+0.017}_{-0.020}$ \\
240.01 - 241.03 &  23.360 $^{+0.021}_{-2.767}$ 	& 488.7 $^{+10.4}_{-5.4}$ 	& -1.412 $^{+0.012}_{-0.013}$ 	& -2.653 $^{+0.028}_{-0.034}$ \\
241.03 - 242.05 &  24.658 $^{+0.261}_{-0.248}$ 	& 250.7 $^{+2.8}_{-2.9}$ 	& -1.294 $^{+0.017}_{-0.017}$ 	& -2.475 $^{+0.018}_{-0.018}$ \\
242.05 - 243.08 &  19.602 $^{+0.203}_{-0.204}$ 	& 280.4 $^{+3.4}_{-3.5}$ 	& -1.298 $^{+0.015}_{-0.015}$ 	& -2.463 $^{+0.020}_{-0.020}$ \\
243.08 - 244.10 &  16.156 $^{+0.173}_{-0.163}$ 	& 280.7 $^{+3.7}_{-3.7}$ 	& -1.327 $^{+0.013}_{-0.013}$ 	& -2.432 $^{+0.020}_{-0.020}$ \\
244.10 - 245.13 &  11.179 $^{+0.113}_{-0.112}$ 	& 284.3 $^{+4.3}_{-4.1}$ 	& -1.355 $^{+0.011}_{-0.011}$ 	& -2.429 $^{+0.020}_{-0.026}$ \\
245.13 - 246.15 &  6.425 $^{+0.062}_{-0.061}$ 	& 285.1 $^{+5.6}_{-5.6}$ 	& -1.473 $^{+0.009}_{-0.009}$ 	& -2.453 $^{+0.034}_{-0.032}$ \\
246.15 - 247.17 &  4.906 $^{+0.047}_{-0.045}$ 	& 316.1 $^{+6.9}_{-7.5}$ 	& -1.498 $^{+0.009}_{-0.009}$ 	& -2.455 $^{+0.043}_{-0.037}$ \\
247.17 - 248.20 &  4.242 $^{+0.040}_{-0.040}$ 	& 309.2 $^{+7.2}_{-7.5}$ 	& -1.500 $^{+0.009}_{-0.009}$ 	& -2.453 $^{+0.042}_{-0.042}$ \\
248.20 - 249.22 &  3.755 $^{+0.036}_{-0.036}$ 	& 328.9 $^{+8.5}_{-8.7}$ 	& -1.521 $^{+0.009}_{-0.009}$ 	& -2.457 $^{+0.047}_{-0.047}$ \\
249.22 - 250.25 &  3.658 $^{+0.035}_{-0.035}$ 	& 299.8 $^{+7.2}_{-7.6}$ 	& -1.503 $^{+0.009}_{-0.009}$ 	& -2.493 $^{+0.052}_{-0.047}$ \\
250.25 - 251.27 &  3.301 $^{+0.032}_{-0.032}$ 	& 283.1 $^{+7.2}_{-7.4}$ 	& -1.514 $^{+0.009}_{-0.009}$ 	& -2.502 $^{+0.055}_{-0.051}$ \\
251.27 - 252.29 &  3.104 $^{+0.031}_{-0.030}$ 	& 276.3 $^{+7.3}_{-7.6}$ 	& -1.544 $^{+0.009}_{-0.009}$ 	& -2.547 $^{+0.065}_{-0.058}$ \\
252.29 - 253.32 &  2.732 $^{+0.027}_{-0.028}$ 	& 265.1 $^{+8.0}_{-8.7}$ 	& -1.597 $^{+0.009}_{-0.009}$ 	& -2.486 $^{+0.063}_{-0.056}$ \\
253.32 - 254.34 &  5.358 $^{+0.050}_{-0.049}$ 	& 459.6 $^{+9.8}_{-10.2}$ 	& -1.421 $^{+0.009}_{-0.009}$ 	& -2.446 $^{+0.041}_{-0.039}$ \\
254.34 - 255.37 &  8.510 $^{+0.081}_{-0.083}$ 	& 502.5 $^{+9.0}_{-8.8}$ 	& -1.348 $^{+0.009}_{-0.009}$ 	& -2.529 $^{+0.038}_{-0.036}$ \\
255.37 - 256.39 &  13.217 $^{+0.115}_{-0.162}$ 	& 460.2 $^{+7.0}_{-7.4}$ 	& -1.326 $^{+0.011}_{-0.010}$ 	& -2.483 $^{+0.026}_{-0.029}$ \\
256.39 - 257.41 &  28.492 $^{+0.214}_{-0.412}$ 	& 703.9 $^{+12.4}_{-13.4}$ 	& -1.391 $^{+0.008}_{-0.012}$ 	& -2.272 $^{+0.018}_{-0.015}$ \\
257.41 - 258.44 &  25.142 $^{+0.269}_{-0.278}$ 	& 760.4 $^{+7.0}_{-8.1}$ 	& -0.783 $^{+0.011}_{-0.009}$ 	& -2.254 $^{+0.012}_{-0.014}$ \\
258.44 - 259.46 &  94.006 $^{+6.131}_{-0.366}$ 	& 272.2 $^{+2.0}_{-1.4}$ 	& 0.754 $^{+0.140}_{-0.016}$ 	& -2.200 $^{+0.009}_{-0.007}$ \\
259.46 - 260.49 &  25.301 $^{+0.293}_{-0.269}$ 	& 651.3 $^{+5.4}_{-5.1}$ 	& -0.610 $^{+0.012}_{-0.012}$ 	& -2.417 $^{+0.014}_{-0.015}$ \\
260.49 - 261.51 &  145.789 $^{+0.825}_{-0.611}$ 	& 260.2 $^{+0.7}_{-1.0}$ 	& 0.018 $^{+0.017}_{-0.009}$ 	& -2.331 $^{+0.010}_{-0.008}$ \\
261.51 - 262.53 &  18.561 $^{+0.227}_{-0.204}$ 	& 798.9 $^{+5.8}_{-6.3}$ 	& -0.448 $^{+0.010}_{-0.010}$ 	& -2.406 $^{+0.014}_{-0.015}$ \\
262.53 - 263.56 &  18.282 $^{+0.662}_{-0.071}$ 	& 833.3 $^{+6.6}_{-6.0}$ 	& -0.430 $^{+0.009}_{-0.010}$ 	& -2.371 $^{+0.015}_{-0.012}$ \\
263.56 - 264.58 &  21.696 $^{+0.250}_{-0.251}$ 	& 751.3 $^{+5.7}_{-6.0}$ 	& -0.539 $^{+0.011}_{-0.010}$ 	& -2.444 $^{+0.014}_{-0.016}$ \\
264.58 - 265.61 &  34.848 $^{+0.427}_{-0.333}$ 	& 529.9 $^{+7.6}_{-5.1}$ 	& -1.146 $^{+0.014}_{-0.012}$ 	& -2.385 $^{+0.015}_{-0.016}$ \\
265.61 - 266.63 &  26.254 $^{+0.213}_{-0.320}$ 	& 643.1 $^{+28.4}_{-11.4}$ 	& -1.754 $^{+0.011}_{-0.009}$ 	& -4.674 $^{+0.024}_{-2.405}$ \\
266.63 - 267.65 &  26.320 $^{+0.274}_{-0.262}$ 	& 256.2 $^{+3.7}_{-3.6}$ 	& -1.426 $^{+0.018}_{-0.019}$ 	& -2.313 $^{+0.013}_{-0.014}$ \\
267.65 - 268.68 &  16.122 $^{+0.170}_{-0.163}$ 	& 332.6 $^{+5.4}_{-5.3}$ 	& -1.380 $^{+0.011}_{-0.014}$ 	& -2.249 $^{+0.016}_{-0.015}$ \\
268.68 - 269.70 &  9.750 $^{+0.098}_{-0.095}$ 	& 423.3 $^{+8.5}_{-8.4}$ 	& -1.413 $^{+0.010}_{-0.009}$ 	& -2.225 $^{+0.019}_{-0.020}$ \\
269.70 - 270.73 &  5.446 $^{+0.050}_{-0.051}$ 	& 458.6 $^{+11.6}_{-12.0}$ 	& -1.475 $^{+0.009}_{-0.009}$ 	& -2.225 $^{+0.026}_{-0.026}$ \\
270.73 - 271.75 &  4.651 $^{+0.042}_{-0.044}$ 	& 446.3 $^{+11.9}_{-12.7}$ 	& -1.491 $^{+0.009}_{-0.009}$ 	& -2.221 $^{+0.028}_{-0.028}$ \\
271.75 - 272.77 &  4.348 $^{+0.040}_{-0.041}$ 	& 425.3 $^{+11.6}_{-12.0}$ 	& -1.500 $^{+0.009}_{-0.009}$ 	& -2.248 $^{+0.030}_{-0.029}$ \\
272.77 - 273.80 &  3.477 $^{+0.032}_{-0.033}$ 	& 497.8 $^{+16.3}_{-16.5}$ 	& -1.526 $^{+0.009}_{-0.009}$ 	& -2.252 $^{+0.038}_{-0.035}$ \\
273.80 - 274.82 &  2.864 $^{+0.027}_{-0.028}$ 	& 558.1 $^{+21.0}_{-21.1}$ 	& -1.539 $^{+0.009}_{-0.009}$ 	& -2.265 $^{+0.044}_{-0.042}$ \\
274.82 - 275.85 &  2.694 $^{+0.026}_{-0.026}$ 	& 566.7 $^{+21.3}_{-21.7}$ 	& -1.526 $^{+0.009}_{-0.009}$ 	& -2.246 $^{+0.045}_{-0.041}$ \\
275.85 - 276.87 &  2.541 $^{+0.025}_{-0.024}$ 	& 538.4 $^{+21.0}_{-20.3}$ 	& -1.536 $^{+0.010}_{-0.009}$ 	& -2.286 $^{+0.055}_{-0.042}$ \\
276.87 - 277.89 &  2.301 $^{+0.024}_{-0.022}$ 	& 544.2 $^{+22.9}_{-22.4}$ 	& -1.554 $^{+0.009}_{-0.010}$ 	& -2.292 $^{+0.053}_{-0.051}$ \\
507.28 - 508.30 &  2.312 $^{+0.024}_{-0.025}$ 	& 224.0 $^{+6.1}_{-6.1}$ 	& -1.474 $^{+0.010}_{-0.010}$ 	& -2.390 $^{+0.047}_{-0.044}$ \\
508.30 - 509.32 &  4.269 $^{+0.040}_{-0.042}$ 	& 230.3 $^{+4.0}_{-4.2}$ 	& -1.300 $^{+0.010}_{-0.009}$ 	& -2.339 $^{+0.027}_{-0.027}$ \\
509.32 - 510.35 &  3.737 $^{+0.068}_{-0.018}$ 	& 584.5 $^{+61.7}_{-2.1}$ 	& -1.494 $^{+0.004}_{-0.005}$ 	& -5.000 $^{+0.000}_{-0.007}$ \\
510.35 - 511.37 &  13.812 $^{+0.130}_{-0.149}$ 	& 255.7 $^{+3.0}_{-3.2}$ 	& -1.161 $^{+0.012}_{-0.012}$ 	& -2.242 $^{+0.014}_{-0.015}$ \\
511.37 - 512.40 &  6.377 $^{+0.058}_{-0.060}$ 	& 293.7 $^{+4.5}_{-4.8}$ 	& -1.267 $^{+0.010}_{-0.010}$ 	& -2.294 $^{+0.022}_{-0.022}$ \\
512.40 - 513.42 &  3.469 $^{+0.033}_{-0.033}$ 	& 317.3 $^{+6.9}_{-7.1}$ 	& -1.379 $^{+0.010}_{-0.010}$ 	& -2.348 $^{+0.037}_{-0.034}$ \\
513.42 - 514.44 &  2.342 $^{+0.024}_{-0.024}$ 	& 312.8 $^{+8.6}_{-9.1}$ 	& -1.431 $^{+0.010}_{-0.010}$ 	& -2.279 $^{+0.039}_{-0.037}$ \\

\enddata
\end{deluxetable}

\end{document}